\documentstyle[12pt,epsfig]{article}

\newlength{\dinwidth}                                                    
\newlength{\dinmargin}                                                    
\def\lapproxeq{\lower .7ex\hbox{$\;\stackrel{\textstyle                                             <}{\sim}\;$}}                                                    
\def\gapproxeq{\lower .7ex\hbox{$\;\stackrel{\textstyle                                                    
>}{\sim}\;$}}                                                    
\def\be{\begin{equation}}                                                    
\def\ee{\end{equation}}                                                    
\def\bea{\begin{eqnarray}}                                                    
\def\eea{\end{eqnarray}}                                                    
\def\bb{b\bar{b}}

\begin{document}                                                    
\titlepage                                                    
\begin{flushright}                                                    
IPPP/08/83   \\
DCPT/08/166 \\                                                    
\today \\                                                    
\end{flushright}                                                    
                                                    
\vspace*{2cm}                                                    
                                                    
\begin{center}                                                    
{\Large \bf Diffractive processes at the LHC\footnote{Based on a talk by A.D. Martin
at Diffraction 2008, La Londe-les-Maures, France, September 2008.} }                                                    
                                                    
\vspace*{1cm}                                                    
A.D. Martin$^a$ ,V.A. Khoze$^{a,b}$ and M.G. Ryskin$^{a,b}$ \\                                                    
                                                   
\vspace*{0.5cm}                                                    
$^a$ Institute for Particle Physics Phenomenology, University of Durham, Durham, DH1 3LE \\                                                   
$^b$ Petersburg Nuclear Physics Institute, Gatchina, St.~Petersburg, 188300, Russia            
\end{center}                                                    
                                                    
\vspace*{0.5cm}                                                    

\begin{abstract}
We present a model of high energy soft $pp$ interactions that has multi $s$- and $t$-channel components, which has been tuned to describe all the available data. The $t$-channel components allow matching of the soft to the hard (QCD) Pomeron. Absorptive effects are found to be large, and, for example, suppress the prediction of the total $pp$ cross section to about 90 mb at the LHC. We use the model to calculate the survival probability, $S^2$, of the rapidity gaps in the exclusive process $pp \to p+H+p$, a process with great advantages for searching for the $H \to \bb$ signal. We consider both eikonal and enhanced rescattering.
\end{abstract}

We discuss two general topics. One is an attempt to obtain a self-consistent description of the high energy behaviour of all soft observables, such as $\sigma_{\rm tot},~ d\sigma_{\rm el}/dt,~ d\sigma_{\rm SD}/dtdM^2,$ particle multiplicities etc., in terms of the underlying physics; and to predict these observables at the LHC. The second topic is to make reliable estimates for the rates of exclusive processes, such as $pp \to p+A+p$, at the LHC, where $A$ is a heavy object and the $+$ signs denote rapidity gaps. Particularly topical is when $A$ is a Higgs boson which decays into $b{\bar b}$; the mass of $A$ can be measured in the exclusive process with very good accuracy
($\Delta M_A\sim 1-2$ GeV) by the missing-mass method by detecting the
outgoing forward protons.  Moreover, a specific $J_z=0$ selection
rule \cite{KMRmm} significantly reduces the $b{\bar b}$ background and also greatly simplifies the spin-parity analysis of $A$. The two topics are inter-related since soft rescattering can destroy the rapidity gaps and strongly deplete the signal. Thus we need a reliable model of soft interactions to estimate the small survival factor $S^2$ of the rapidity gaps.

\section{Soft scattering including absorptive effects}
The total and elastic proton-proton cross sections
  are usually described in terms of an eikonal model, which automatically
satisfies  $s$-channel
elastic unitarity. To account for the possibility of excitation of the initial proton, 
that is for two-particle intermediate states with the
proton replaced by $N^*$, we use the Good-Walker
formalism \cite{GW}. Already at Tevatron energies the absorptive
correction to the elastic amplitude, due to elastic eikonal
rescattering, gives about a 20\% reduction of
simple one Pomeron exchange. After accounting for low-mass proton
excitations (that is $N^*$'s in the intermediate states) the correction becomes twice
larger (that is, up to 40\%). Next, in order to describe high-mass diffractive
dissociation, $d\sigma_{\rm SD}/dM^2$, we have to include an extra factor
of 2  from the AGK cutting rules \cite{AGK}. Thus, the absorptive effects
in the triple-Regge domain are expected to be quite large. The previous
triple-Regge analyses (see, for example, \cite{FF}) did not allow for
absorptive corrections and the resulting triple-Regge couplings must be
regarded, not as bare vertices, but as effective couplings
embodying the absorptive effects. 
Thus, we perform a {\it new triple-Regge analysis} of the fixed-target FNAL, CERN-ISR and Tevatron data
that includes the absorptive effects explicitly \cite{LKMR}.
The bare triple-Pomeron coupling is found to be about three times larger than before; now $g_{3P}=\lambda g_n$ with $\lambda \simeq 0.2$ where $g_N$ is the Pomeron-proton coupling.

\subsection{Inclusion of multi-Pomeron vertices}
\begin{figure}
\begin{center}
\includegraphics[height=3cm]{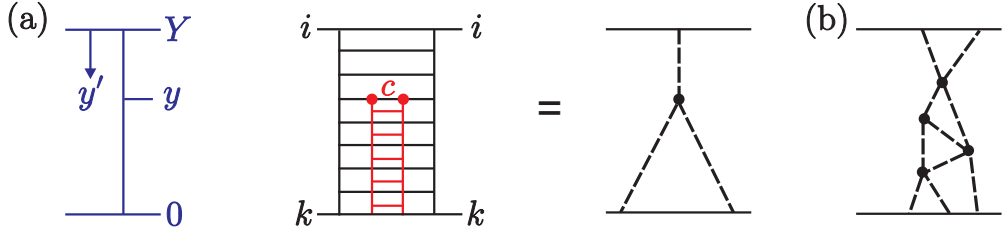}
\caption{(a) The ladder structure of the triple-Pomeron amplitude between diffractive eigenstates $i,k$ of the proton; the rapidity $y$ spans an interval 0 to Y={\rm ln}s. (b) A multi-Pomeron diagram.} 
\label{fig:evol}
\end{center}
\end{figure}

Since the triple-Pomeron vertex turns out to be rather large,
the contribution of the so-called `enhanced' diagrams, with a few
vertices, is not negligible. Moreover, it is more reasonable to include multi-Pomeron contributions with $n$ to $m$ Pomeron vertices given by $g^n_m\propto
\lambda^{n+m}$ than to assume that $g^n_m=0$ for $n+m>3$.  These effects are included in the following evolution equation for the opacity $\Omega$ in rapidity space, at given impact parameter $b$
\begin{equation}
\frac{d\Omega_k(y,b)}{dy}\,=\,e^{-\lambda\Omega_i(y',b)/2}~~~e^{-\lambda
\Omega_k(y,b)/2}~~
\left(\Delta+\alpha'\frac{d^2}{d^2b}\right)\Omega_k(y,b)\; ,
\label{eq:evol1}
\end{equation}
$$~~~~~~~~~~~~~~<-({\rm c})->~~<-({\rm b})->~~<---({\rm a})--->$$
where $y'=\ln s -y$.   Let us explain the meanings of the three terms on the right hand side of (\ref{eq:evol1}). If only the last factor, (a), is present then the evolution generates the ladder-type structure of the bare Pomeron exchange amplitude, where the Pomeron trajectory $\alpha_P=1+\Delta+\alpha't$. The inclusion of (b) allows for rescatterings of an intermediate parton $c$ with the ``target'' proton $k$; Fig.~\ref{fig:evol}(a) shows the simplest (single) rescattering which generates the triple-Pomeron diagram. Finally, (c) allows for rescatterings with the beam $i$. In this way we include the absorptive effects generated by all multi-Pomeron diagrams like the one shown in Fig.~\ref{fig:evol}(b). 
There is an analogous equation for the evolution of $\Omega_i(y,b)$, and the two equations may be solved iteratively. A detailed discussion can be found in \cite{KMRs1},
where $\alpha'$ is set to zero.

\subsection{Inclusion of different $t$-channel components}
Besides allowing for $\alpha'\neq 0$, a new development is an analysis \cite{KMRs2} in which we allow for four different $t$-channel states, which we label $a$: one for the
secondary Reggeon ($R$) trajectory and three Pomeron states ($P_1, P_2, P_3$) to mimic the BFKL
diffusion in the logarithm of parton transverse momentum,
$\ln(k_t)$ \cite{Lip}. To be precise, since the
BFKL Pomeron \cite{bfkl} is not a pole in the complex $j$-plane, but a
branch cut,
we approximate the cut by three  $t$-channel states of a different size.
The typical values of $k_t$ in each of the three states is about $k_{t1}\sim 0.5$ GeV,
$k_{t2}\sim 1.5$ GeV and $k_{t3}\sim 5$ GeV.
Thus (\ref{eq:evol1}) is replaced by
\begin{equation}
\frac{d\Omega^a_k(y,b)}{dy}\,=\,
e^{-\lambda\Omega_i(y',b)/2}~e^{-\lambda \Omega_k(y,b)/2}~
\left(\Delta^a+\alpha'\frac{d^2}{d^2b}\right)\Omega^a_k(y,b)
+V_{aa'}\Omega^{a'}_k\; .
\label{eq:evol2}
\end{equation}
where $\Delta^a=\alpha_a(0)-1$ and $\alpha'_a=\alpha'_P$ for $a=P_1,P_2,P_3$, while for the secondary Reggeon, $(a=R)$, which is built of quarks, we take $\Delta^R=\alpha_R(0)=0.6$ and $\alpha'_R=0.9~ {\rm GeV}^{-2}$. The transition factors $V_{aa'}$ were fixed by properties of the BFKL equation.
In the exponents, the opacities $\Omega_i$
($\Omega_k$) are actually the sum of the opacities $\Omega^{a'}_i$
($\Omega^{a'}_k$) with corresponding coefficients. 

\begin{figure}
\begin{center}
\includegraphics[height=15cm]{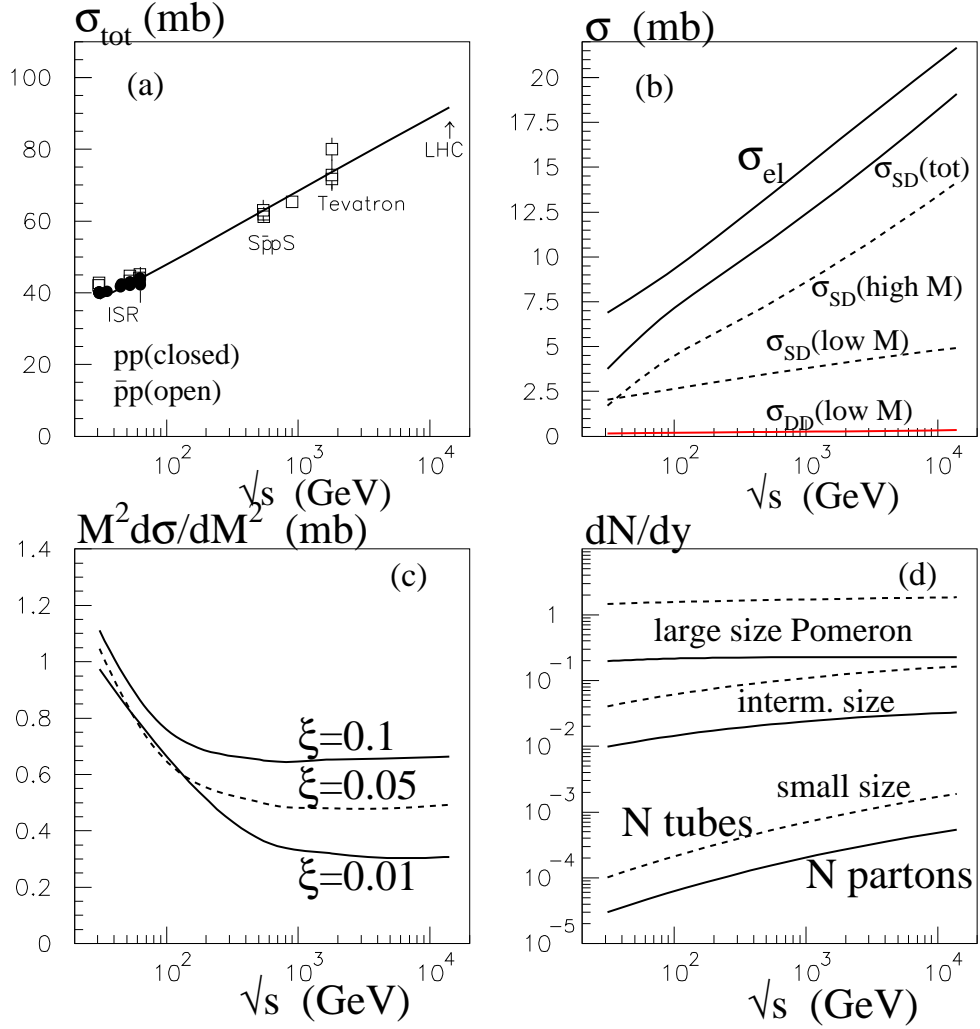}
\caption{The energy dependence of (a) the total, (b) the elastic and
diffractive dissociation, $pp$ cross sections and (c) the cross sections of
dissociation to a fixed $M^2$ state, where $\xi=M^2/s$. Plot (d) shows the parton
multiplicity (solid lines) and the number of `colour  tubes' (dashed)
produced by the Pomeron components of different size.}
\end{center}
\label{fig:data}
\end{figure}
Such a model allows us to reproduce all the available data on
diffractive cross sections, $\sigma_{\rm tot},\, d\sigma_{\rm el}/dt,\,
\sigma_{\rm SD}^{\rm low\, mass},\,
d\sigma_{\rm SD}/dM^2$. We find that the triple-Pomeron coupling parameter $\lambda=0.25$ and that $\Delta^a=0.3$ for all three components of the Pomeron, consistent with the expectations of resummed NLL BFKL which gives $\omega_0\equiv\alpha_P(0)-1 \sim 0.3$ pratically independent of $k_t$ \cite{BFKLnnl}. The slopes, $\alpha'_P$, are 0.05, 0.005 and 0 GeV$^{-2}$ for the large, intermediate and small size components of the Pomeron respectively.

The model allows us, in principle, to predict all features of soft $pp$ high energy interactions. Examples are shown
in Fig.~\ref{fig:data}. Note that, due to absorptive effects, the total cross section at
the LHC energy $\sqrt{s}=14$ TeV is predicted to be only about 90 mb. Also, the
multiplicities of the secondaries produced by the $t$-channel Pomeron
components of different sizes, are shown in Fig.~\ref{fig:data}(d).
We see that, starting with the same `bare' intercepts ($\Delta=0.3$), after the absorptive correction, the contribution of the large-size
component becomes practically
flat, while  the small-size contribution, which is much less affected by
the absorption, continues to grow with energy. Such a behaviour is
consistent with  experiment, where the density of low $k_t$
secondaries is practically saturated, while the probability to produce a hadron
with a large transverse momentum (say, more than 5 GeV) grows with the
initial energy.

\section{Exclusive Processes}

\begin{figure}
\begin{center}
\includegraphics[height=4cm]{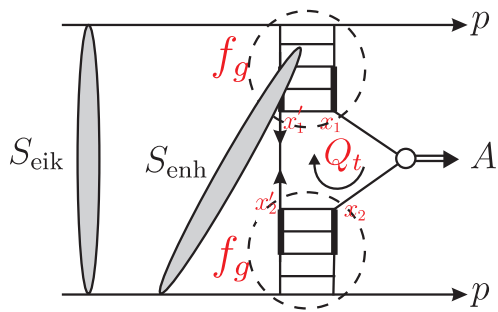}
\caption{The mechanism for the exclusive process $pp \to p+A+p$, with the eikonal and enhanced survival factors shown symbolically. The thick lines on the Pomeron ladders, either side of the subprocess ($gg \to A$), indicate the rapidity interval $\Delta y$ where enhanced absorption is not permitted \cite{Dy}.}
\label{fig:pAp}
\end{center}
\end{figure}

We have already emphasised the value of the observation of an exclusive process of the type $pp \to p+A+p$ at the LHC \cite{KMRProsp,DKMOR,FP420}.
 The process is sketched in Fig.~\ref{fig:pAp}. The case of $A=H \to \bb$ is particularly interesting. The cross section is usually written in the form
\be
\sigma (pp \to p+A+p) ~\sim~\frac{\langle S^2 \rangle}{B^2} \left| N\int \frac{dQ_t^2}{Q^4_t}f_g(x_1,x'_1,Q^2_t,\mu^2)f_g(x_2,x'_2,Q^2_t,\mu^2) \right| ^2
\label{eq:d1}
\ee
where $B/2$ is the $t$-slope of the proton-Pomeron vertex, and the constant $N$ is known in terms of the $A \to gg$ decay width. The amplitude-squared factor, $\mid...\mid^2$, can be calculated in perturbative QCD, since the dominant contribution to the integral comes from the region $\Lambda^2_{QCD} \ll Q^2_t \ll M^2_A$, for the large values of $M_A^2$ of interest. The probability amplitudes, $f_g$, to find the appropriate pairs of $t$-channel gluons $(x_1,x'_1)$ and $(x_2,x'_2)$ of Fig.~\ref{fig:pAp}, are given by skewed unintegrated gluon densities at a hard scale $\mu \sim M_A/2$.
To evaluate the cross section of such an exclusive processes it is
important to know the probability, $\langle S^2 \rangle$, that the rapidity gaps survive
and will not be filled by secondaries from eikonal and enhanced
rescattering effects.

The model of soft interactions described above gives a gap
survival probability $\langle S^2_{\rm eik} \rangle \sim 0.012$\footnote{If we were to adjust $\langle S^2_{\rm eik} \rangle$, obtained with the $t$ dependence of our model, to its value corresponding to an exponential slope $B=4$ GeV$^{-2}$, then we obtain $\langle S^2_{\rm eik} \rangle_{\rm eff}=0.024$, which is consistent with our previous estimate \cite{KMRsoft}, see the discussion below.} with respect to the {\it eikonal}
(including the elastic and low-mass proton excitation) rescattering, for the exclusive production of a Higgs boson. The model may also be used to calculate the absorptive correction to exclusive
cross sections caused by the so-called {\it enhanced} diagrams, that is by
the interaction with the intermediate partons, see Fig.~\ref{fig:pAp}. This rescattering
violates `soft-hard' factorisation, since the probability of
such an interaction depends both on the transverse momentum and on the
impact parameter of the intermediate parton. The model predicts 
$\langle S^2_{\rm enh}\rangle \sim \frac{1}{3}$. 

We emphasize that comparing the values of the survival factors in this way is too simplistic. The problem is that, with enhanced screening on intermediate partons, we no longer have exact factorisation between the hard and soft parts of the process. Thus, before computing the effect of soft absorption we must fix what is included in the bare exclusive amplitude calculated in terms of perturbative QCD. Two observations are importrant.

The {\it first} observation is that the bare amplitude is calculated as a convolution of two generalised (skewed) gluon distributions with the hard subprocess matrix element, see (\ref{eq:d1}). These gluon distributions are determined from integrated gluon distributions of a global parton analysis of mainly deep inelastic scattering data. Now, the phenomenological integrated parton distributions already include the interactions of the intermediate partons with the parent proton. Thus calculations of $S_{\rm enh}$ should keep only contributions which embrace the hard matrix element of the type shown in Fig.~\ref{fig:pAp}.

The {\it second} observation is that the phenomenologically determined generalised gluon distributions, $f_g$, are usually taken at $p_t=0$ and then the observed ``total'' exclusive cross section is calculated by integrating over $p_t$ of the recoil protons assuming the an exponential behaviour $e^{-Bp_t^2}$; that is
\be
\int dp^2_t~e^{-Bp_t^2}~=~1/B~=~\langle p_t^2\rangle.
\ee
However, the total soft absorptive effect changes the $p_t$ distribution in comparison to that for the bare cross section determined from perturbative QCD. Thus the additional factor introduced by the soft interactions is not just the gap survival $S^2$, but rather the factor $S^2/B^2$ \cite{KMRProsp,KMRxc}, which strictly speaking has the form  $S^2\langle p^2_t \rangle^2$. 

In order to compare determinations of the suppression due to absorptive effects we should compare only the values of the complete cross section for $pp \to p+A+p$. However, a comparison is usually made by reducing the cross section to a factorized form. If this is done, as in (\ref{eq:d1}), then the predictions for the survival factor to eikonal and enhanced screening of the exclusive production of a 120 GeV Higgs at the LHC are $\langle S^2 \rangle_{\rm eff}=0.004,~0.009,~ 0.015$ where enhanced screening is only permitted outside a threshold rapidity gap \cite{Dy} $\Delta y=0,~1.5,~2.3$ respectively, see Fig.~\ref{fig:pAp}. The values correspond to $B=4~{\rm GeV}^{-2}$. That is, allowing for the threshold effect, we predict $\langle S^2 \rangle_{\rm eff} \simeq 0.015 \pm 0.01$.

\section{Summary}

We have described a model, tuned to the existing data, which is capable of predicting all features of high energy soft $pp$ interactions. Absorptive effects are found to be large. For example, the total $pp$ cross section is predicted to be only about 90 mb at the LHC energy $\sqrt{s}=14$ TeV. We used this model to estimate the exclusive cross section for Higgs production, $pp \to p+H+p$, at the LHC. We calculated the survival factor of the rapidity gaps to both eikonal and enhanced rescattering and found $\langle S^2 \rangle_{\rm eff} \simeq 0.015 \pm 0.01$. Note that, from exclusive CDF data and leading neutron data at HERA, there is evidence that $\langle S^2_{\rm enh} \rangle$ is somewhat larger than the estimates obtained here, such that $\langle S^2 \rangle_{\rm eff}$ is nearer the upper limit of the quoted interval. Early LHC runs can measure $\langle S^2_{\rm enh} \rangle$ \cite{KMRearly}.

\end{document}